\documentclass[A4paper,preprint,aps,amsmath,10pt,twocolumn]{revtex4-1}
\usepackage{epsfig}

\begin{document}

\title{Quantitative study of electronic whispering gallery modes in electrostatic-potential induced circular graphene junctions}
\author{T Lien Le$^1$ and V Lien Nguyen$^{2,3}$ }
\email{nvlien@iop.vast.ac.vn}
\affiliation{$^1$University of Science and Technology of Hanoi, Hanoi, Vietnam \\
$^2$ Institute for Bio-Medical Physics, 109A Pasteur, 710115 Hochiminh City, Vietnam \\
$^3$Theoretical and Computational Physics Dept., Institute of Physics, VAST \\ 
10 Dao Tan, Ba Dinh Distr., 118011 Hanoi, Vietnam } 
 
\date{15 September 2019}

\vspace*{1cm}

\begin{abstract}
Electronic Whispering Gallery Modes (EWGMs) have been recently observed in several circular graphene junctions, $pn$ and $pp'$, created in scanning tunnelling microscope experiments. By computing the local density of states within the Dirac-Weyl formalism for massless fermions we demonstrate that the EWGMs may really be emerged in any type of the electrostatic-potential induced circular graphene junctions, including uni-junctions (e.g. $np$- or $pp'$-junctions)  as well as bipolar-junctions (e.g. $pnp$-heterojunctions). Surprisingly, quantitative analyses show that for all the EWGMs identified (regardless of junction types) the quality ($Q$) factors seem to be $\leq 10^2 $, very small compared to those in ordinary optical whispering gallery modes microresonators, while the corresponding mode radii may tunably be in nanometer-scale. Our theoretical results are in good agreement with existent experimental data, putting a question to the application potential of the EWGMs identified.
\end{abstract}

\maketitle
\section{Introduction}
The optical microresonators (or microcavities) that confine the light to small volumes by resonant recirculation are widely utilised in modern linear and nonlinear optics \cite{vahala2003}. The most desirable resonators would confine light without loss and would have resonant frequencies defined precisely. In practice, optical resonators are characterised by the two parameters, the quality factor ($Q$-factor) and the mode volume ($V$), that respectively describe the temporal and spatial confinement of light in devices. Resonators with potential applications are those of high $Q$ and small $V$. It appears that an extremely high value of $Q$ may be achieved in the so-called whispering-gallery microresonators of very small volume \cite{oraevsky2002, matsko2006, pollinger2009, acharyya2019}. In these microresonators, like dielectric microspheres, microdisks, or microtori, the light is effectively confined by repeated total internal reflections at the curved boundaries, giving rise to resonances. The circular optical modes emerged in such resonators are often referred to as Whispering-Gallery Modes (WGMs). The $Q$-factor of optical WGMs may be as high as $\sim 10^{10}$, depending primarily on the resonator material and a perfection of dielectric surfaces \cite{matsko2006}. With a very high $Q$ in combination of other advantages such as very small mode volume and very simple geometry-structure, WGM-resonators emerged as the most potential optical resonators for a variety of applications \cite{oraevsky2002, matsko2006}.

As well-known, there is a close similarity between light-rays in geometrical optics and ballitic trajectors of electrons. This similarity attracted much more attention by the discovery of graphene, in which massless charge-carriers exhibit the photon-like linear dispersion and gain a very large mean-free path (of micrometer even at room temperature) \cite{castro2009}. It was established that the transport of electrons through an electrostatic potential barrier in a graphene heterostructure may well resemble the optical refraction at a surface of metamaterials with negative refractive index \cite{cheianov2007, ghlee2015}. As a consequence, the graphene $np$-junctions could be perfectly used to create an electronic analogue of Veselago optical lens \cite{cheianov2007}. And, moreover, the scanning tunnelling microscope (STM)-tip induced circular graphene $np$-junctions that are extensively exploited to study different properties of Dirac fermions confined by an axially symmetric electrostatic potential barrier \cite{gutierrez2016, jlee2016, freitag2016} should act as electronic WGM-resonators in producing circular electronic modes analogous to the optical WGMs. Indeed, recently, electronic Whispering-Gallery Modes (EWGMs) have been reported in several STM-experiments  \cite{zhao2015, ghahari2017, jiang2017}. Owing to the dual-gate structure, the back-gate and top-gate, STM-based EWGM-resonators are fully tunable in the meaning that both the resonator size and the $np$-interface potential may be independently varied by changing suitably the back-gate voltage, the top-gate potential and the tip-to-graphene distance \cite{jiang2017}. EWGMs in these resonators can be detected by measuring the tunnelling differential conductance that feature the local density of states (LDOSs) spectrum in dependence on the tip-sample bias, back-gate voltage, and spatial position (from the centre of the tip). So, the observed EWGMs can be theoretically understood by calculating the LDOS for the massless Dirac-like fermions under a suitable tip/gate-induced electrostatic potential. In the continuum calculation reported in Ref. \cite{zhao2015} this potential is simply assumed to have the parabolic form, while in the tight-binding model used in Ref.\cite{jiang2017} it is the Thomas-Fermi approximated potential. Both the studies have unambiguously confirmed an emergence of EWGM-spectra in STM-tip induced circular graphene resonators. Here, we note that all the studies in Refs.\cite{zhao2015, ghahari2017, jiang2017} concern the resonators with $np$/$pn$-junctions.Very recently, it was reported that similar EWGMs have been observed even in the STM-tip induced circular graphene resonator with $pp'$-junctions \cite{ren2019}.

Actually, EWGMs are known as an almost periodic sequence of resonances emerged in an energy spectrum of a resonator. For the circular graphene resonators under study, these resonances truly describe the quasi-bound states (QBSs) that are formed as a result of interference processes of the electronic waves, undergone multiple Klein-scatterings by the electrostatic confinement potential on the inside of the resonator \cite{matulis2008}. Generally, QBSs could be created by any electrostatic confinement potentials \cite{matulis2008, chen2007}. The structure of QBS-spectra however depends on the interference pattern of wave functions inside the resonator, and the later, in turn, is highly sensitive to the characteristics of the confinement potential (such as its magnitudes, signs or sizes). Also, these characteristics are closely correlated with each other in affecting the QBS-spectra. So, it seems that to create a QBS-spectrum with EWGMs in an electrostatic-potential induced circular graphene resonator of any junction-type one has just to set the appropriate characteristics to the confinement potential. And, in principle, EWGMs may emerge in any type of  these junctions, though the chance of getting them as well as their quality, i.e. $Q$-factor and mode volume $V$, might be different, depending on the junction type. Since these quantities, $Q$ and $V$, are primary characteristics of EWGMs, one certainly  has to determine them first in examining EWGM-spectra. 
    
The purpose of the present theoretical work is to quantitatively study the EWGMs emerged in various models of circular graphene junctions, including uni-junctions such as $np$-junctions (CGNPJs) or  $pp'$-junctions (CGPP'Js) and bipolar-junctions such as $pnp$- heterojunctions (CGPNPHJs). The junctions are assumed to be created by axially symmetric electrostatic potentials like those in STM-experiments. The study was carried out within the framework the Dirac-Weyl formalism for massless fermions in the presence of the suggested confinement potential. For each of these resonator-models we searched for EWGMs by analysing the LDOSs calculated in wide value-ranges of confinement-potential parameters. For all the identified EWGMs we evaluated the $Q$-factors and the effective mode radii \cite{note01}, following the way that is often used for optical WGM-resonators. Qualitatively, our studies demonstrate that the EWGMs may emerge in electrostatic-potential induced circular graphene resonators with any type of  junctions, depending primarily on the confinement potential parameters. Quantitative analyses show that for all the EWGMs identified the $Q$-factors seem always to be $\leq 10^2 $, very small compared to those in ordinary optical WGM-microresonators (of $\simeq 10^5 - 10^8$ \cite{vahala2003, matsko2006}), while the corresponding mode radii may tunably be in nanometer-scale. Our theoretical results are in a good agreement with existent experimental data \cite{zhao2015, jiang2017, ren2019}, putting  a question to the application potential of the EWGM identified.

Thus, we are interested in the circular graphene junctions created by an axially symmetric electrostatic confinement-potential $U(r)$ in a continuous single-layer graphene sheet. Neglecting the valley scattering and using the units such that $\hbar = 1$ and the Fermi velocity $v_F = 1$, the low-energy electronic excitations in these structures can be described by the two-dimensional (2D) massless Dirac-Weyl Hamiltonian:
\begin{equation}    
{\cal H} = \vec{\sigma} \vec{p} + U(r), 
\end{equation}
where $\vec{\sigma} = (\sigma_x , \sigma_y )$ the Pauli matrices and $\vec{p} = - i (\partial_x , \partial_y )$ the 2D momentum operator. In STM-experiments $U(r)$ is mainly resulted from a combined effect of the tip-sample potential and the back-gate voltage.

\begin{figure}[htb]
\begin{center}
\includegraphics[width=0.5\textwidth]{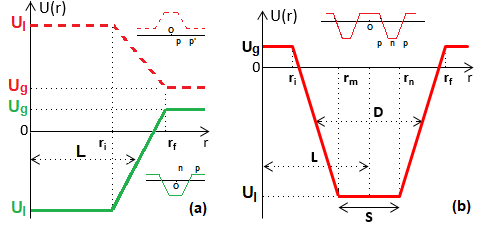}
\end{center}
\caption{\small \textit{(color online) Models of radial electrostatic potentials $U(r)$ that create the two types of circular graphene junctions under study: $(a)$ Uni-junctions: green-solid profile - CGNPJs and  red-dashed profile - CGPP'Js with average radius $L$  and $(b)$ Bipolar-junctions: red-solid profile - CGPNPJs with  average radius $L$ and average width $D$. All the modelled potentials are constant at the limiting distances of $r \leq r_i$ and $r \geq r_f$.}}
\label{fig:1}
\end{figure}

Given $U(r)$, we computed LDOSs for the studied resonator, using the approach suggested in Refs.\cite{chau2016, chau2017}(see Supplementary Materials). Shortly, the computing procedure is as following: $(i)$ solving the Dirac equation of Hamiltonian (1) to calculate LDOSs with a given angular momentum $j$ - the partial LDOSs (S4); $(ii)$ taking the sum of partial LDOSs over all possible $j$ provides LDOS $\rho (E, r)$ (S3) that depends on the energy $E$ and the distance $r$; and $(iii)$ integrating $\rho (E, r)$ over $r$ provides the total density of states (TDOS) $\rho_T (E)$ (S9). All the features of a resonance spectrum are definitely manifested in its LDOS and TDOS. Certainly, no all junction-samples may reveal EWGMs. So, we had to search for these modes, varying different confinement-potential parameters. Qualitatively, EWGMs can be identified as a spectrum with an almost periodic sequence of resonances, appearing in a narrow range of energy in one side and close to the charge neutrality point, while in the other side the spectrum shows itself to be featureless \cite{jiang2017}. 

Once a EWGM-spectrum is identified we have quantitatively examined each of most profound resonances in the spectrum by evaluating its partial quality-factor $Q_i$ and partial mode radius $R_i$ \cite{note01}. To this end, for the resonance at energy $E_i$, we measure the resonance width $\delta E_i$ (by fitting resonance peak into an appropriate Lorentzian profile) and the resonance spacing $\Delta E_i$ (see Fig.6$(d)$). Quantities $Q_i$ and $R_i$ could be then determined in the way as that used for optical WGM-resonators: $Q_i = \omega_i \tau_i \equiv | E_i | / \delta E_i$ and  $R_i \approx \hbar v_F / \Delta E_i$, where $\omega_i = |E_i | / \hbar$ is the resonant-mode frequency and $\tau_i = \hbar / 2 \delta E_i$ is the lifetime of the mode ($\hbar = Planck \ constant / 2\pi $ and $v_F \approx 10^6 m/s$ is the Fermi velocity) \cite{matsko2006,zhao2015,pollinger2009}. From partial quantities $[Q_i ]$, $[R_i ]$ and $[\Delta E_i ]$ we respectively deduced the average quantities $Q$, $R$ and $\Delta $ which could be used to characterise the examined EWGM-spectrum on the whole. Such studies have been realised for all the EWGM-spectra identified in circular graphene resonators with different junctions types, uni-junctions CGNPJs and CGPP'Js as well as bipolar-junctions CGPNPHJs.

Thus, first we have to define radial confinement potentials $U(r)$. For the resonators with uni-junctions, the potential $U(r)$ is chosen in the form (see Fig.1$(a)$):
\begin{equation}
U(r) = \left\{  \begin{array}{ll}  
U_l \ \ {\rm for} \  r  \leq  r_i   \\
U_l + \frac{r - r_i }{r_f - r_i } (U_g - U_l ) \ \ {\rm for} \  r_i < r < r_f  \\   
U_g \ \ {\rm for} \  r \geq r_f  .   
   \end{array}   \right.
\end{equation}  
The distances $r_i$ and $r_f$ in this equation can be merely expressed as $r_i = (1 - \alpha ) L$ and $r_f = (1 + \alpha ) L$, where the quantity $\alpha$ with $0 \leq \alpha \leq 1$ and the length $L$ respectively measure the smoothness of the junction-boundary potential and the average radius of the junction (see Fig.1$(a)$). So, the potential $U(r)$ suggested in eq.(2) is entirely characterised by the four parameters: $U_l$, $U_g$, $L$, and $\alpha$. In relation to the STM-experiments, the potentials $U_g$ and $U_l$ should be thought of as defined respectively by the back-gate voltage and the tip-sample and back-gate voltages combined, while the two other parameters, $L$  and $\alpha$, are essentially related to the tip size and the tip-sample distance \cite{jiang2017}. The potential $U(r)$ of eq.(2) is quite general, describing all possible circular graphene uni-junctions. Particularly, this potential $U(r)$ describes CGNPJs if $U_l < 0$ and $U_g > 0$. In the other case, when both $U_l$ and  $U_g $ are positive, it describes CGPP'Js. Here, it is useful to note that due to the electron-hole symmetry in the models under study a simultaneous change in sign of the two potentials $U_l$ and $U_g$ as well as the energy $E$ does not make the spectrum changed. So, we should consider only two types of uni-junctions, e. g. CGNPJs and CGPP'Js. Certainly, this note should also be applied to the bipolar-junctions introduced below.

In order to model the circular graphene bipolar-junctions, i.e $npn$- or $pnp$-heterojunctions, we define the radial confinement potential $U(r)$ as (see Fig.1$(b)$).
\begin{equation}
U(r) = \left\{  \begin{array}{ll}  
U_g    \ \ {\rm for} \  r  \leq  r_i   \\
U_g + \frac{r - r_i }{r_m - r_i}(U_l - U_g )  \ {\rm for} \ r_i  \leq  r \leq  r_m \\ 
U_l     \ \ {\rm for} \ r_m  <  r  < r_n  \\
U_l - \frac{r - r_n } {r_f - r_n}(U_l - U_g ) \ {\rm for} \  r_n  \leq  r \leq  r_f  \\   
U_g     \ \ {\rm for} \  r \geq  r_f  .   
   \end{array}   \right.
\end{equation}
 Actually, the labelled distances ($r_\nu , \nu = i, m, n, f $) in this expression can be expressed as $r_i = L - D + S/2, \ r_m = L - S/2, \ r_n = L + S/2$ and $r_f = L + D - S/2$, where $L$, $D$, and $S$ may be effectively understood as the average radius of junction, its average width and the tip/top-gate size, respectively (see Fig.1$(b)$). Thus, in the model suggested a  circular graphene bipolar junction is characterised by the five parameters: $U_l$, $U_g$, $L$, $D$, and $S$. The potentials $U_l$ and $U_g$ could be here thought of as having the same source as those in the potential of eq.(2). In addition, to describe bipolar-junctions these two potentials must be different in sign, $U_l . U_g < 0$, implying the two possible cases of sign-realisations. However, as noted above on the electron-hole symmetry, we need consider only one of these cases, e.g. the case of $U_l < 0$ and $U_g > 0$ (i.e. CGPNPHJs in Fig.1$(b)$). Note that an equal smoothness is explicitly introduced at both heterojunction boundaries in the potential $U(r)$ of eq.(3). 

Importantly, both the potentials in eqs.2 and 3 become constant in the two limits of small and large distances, $r \leq r_i$ and $r \geq r_f$, that would somewhat facilitate the LDOS-computations \cite{chau2017}. In particular case of $U_g \equiv 0$, these potentials $U(r)$ of eqs(2) and (3) seem to have the ordinary trapezoidal profiles. The trapezoidal potentials are often used to describe the gate-induced graphene structures \cite{huard2007, sonin2009}, which are also referred to as circular graphene quantum dots \cite{matulis2008, chau2016} or quantum rings \cite{cabosart2017, linh2018}. In reality, a trapezoidal shape is quite a good fit of the Lorentzian shape that is widely believed to be the profile of electrostatic potentials induced by a STM-tip \cite{downing}.  An advantage of the potentials of eqs.(2) and (3) also lies in their simplicity so that the Hamiltonian of eq.(1) could be exactly solved \cite{chau2016, sonin2009}. 

\begin{figure*}[htb]
\begin{center}
\includegraphics[width=0.7\textwidth]{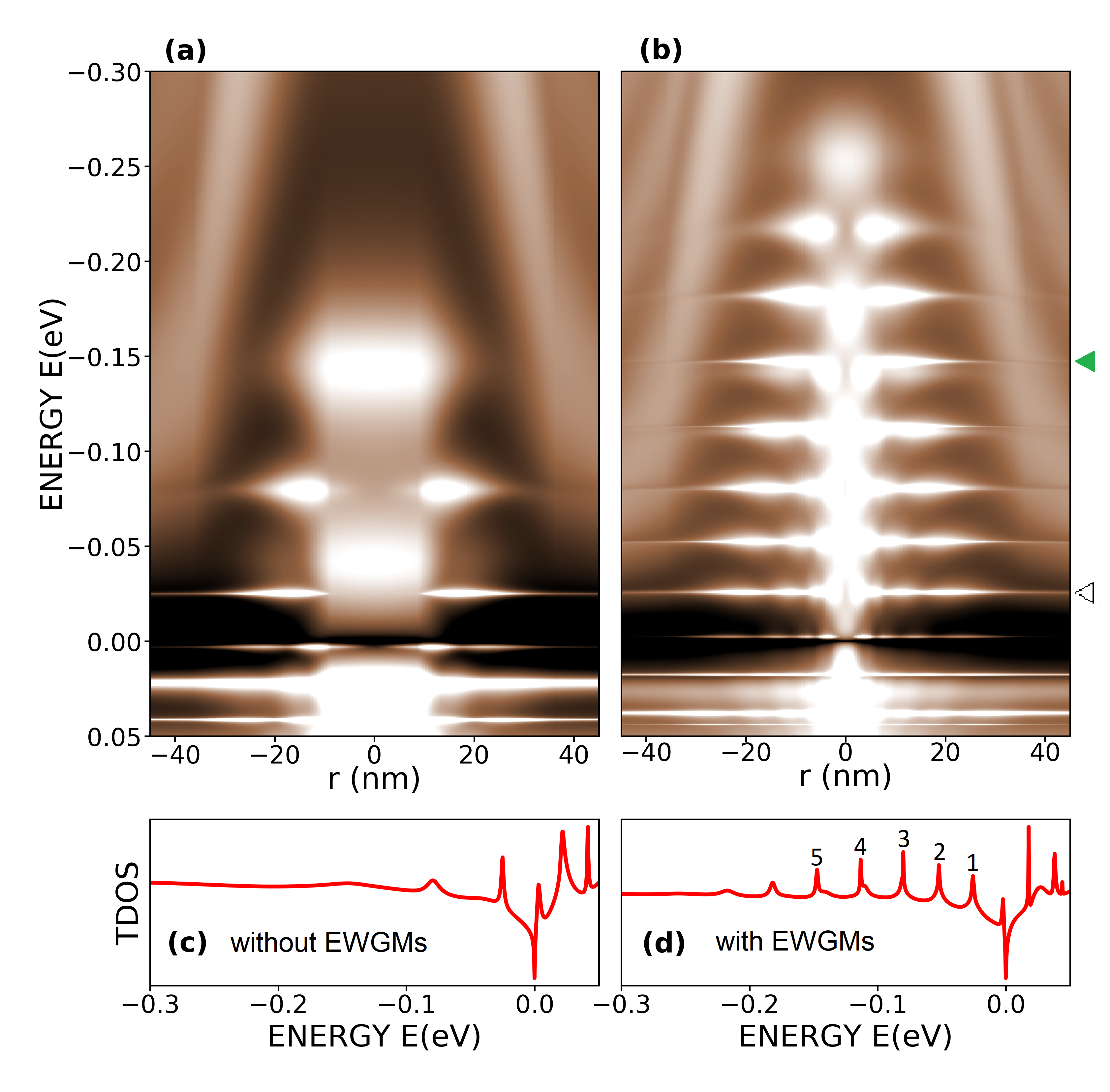}
\end{center}
\caption{\small \textit{(color online) $(a)$ Map of LDOS as a function of distance $r$ and $(c)$ corresponding TDOS (in arbitrary unit) for the CGNPJ with potential parameters $[U_l , U_g , L, \alpha ] = [-0.35 \ eV, 0.02 \ eV, 15 \ nm, 0.9]$. $(b)$ and $(d)$ are the same as $(a)$ and $(c)$, respectively, but for another CGNPJ with $[U_l , U_g , L, \alpha ] = [-0.35 \ eV, 0.02 \ eV, 30 \ nm, 0.5]$. While the spectrum in $(a,c)$ is featureless, that in $(b,d)$ clearly shows EWGMs. }}
\label{fig:2}
\end{figure*}

So, given parameters of the potential $U(r)$ of  eq.(2) or eq.(3), as mentioned above, we solved the eigenvalue equation for the Hamiltonian of eq.(1), computed the LDOSs, searched for EWGM-spectra, and quantitatively analysed the EWGMs identified. Searching for EWGMs requires a bit of patience, though some guesses can be made, using experimental data for uni-junctions (for CGNPJs \cite{jiang2017} and for CGPP'Js \cite{ren2019}). Anyway, we were able undoutedly to identify the EWGMs in resonators with any type of junctions under study. In the case of uni-junctions, identified EWGMs  resemble well existent experimental data. Below, in Figs.2-3, Fig.4, and Fig.5 we present the computational results for the CGNPJs, CGPP'J, and CGPNPHJ, respectively. These figures have the same structure, showing the qualitative behavior and quanlitative characters of the EWGM examined. So, avoiding an unnecessary repeat, most detailed discussions relating to the CGNPJ in Figs.2-3 may also be applied to the CGPP'J in Fig.4 as well as the CGPNPHJ in Fig.5.

Fig.2 presents the computed maps of LDOSs as a function of distance $r$ (boxes $(a)$ and $(b)$) and  corresponding TDOSs (boxes $(c)$ and $(d)$) for the two CGNPJs with different parameter values of the potential of eq.(2) (given in the caption to the figure). Indeed, both the spectra in $(c)$ and $(d)$ show the resonances (or QBSs) which however carry very different features. The spectrum in box $(c)$ is featureless, showing no particular relation between the magnitudes as well as the positions of  emerged resonances (Here, one might think of the so-called atomic collapse resonances \cite{wang2013}). On the contrary, the spectrum in box $(d)$ shows an almost periodic sequence of resonances, appeared on one energy-side from the neutral point. This is the typical feature of EWGMs. To ensure that the LDOS in Fig.2$(b)$ really manifests a EWGM-spectrum we should further explore it.

\begin{figure*}[htb]
\begin{center}
\includegraphics[width=0.8\textwidth]{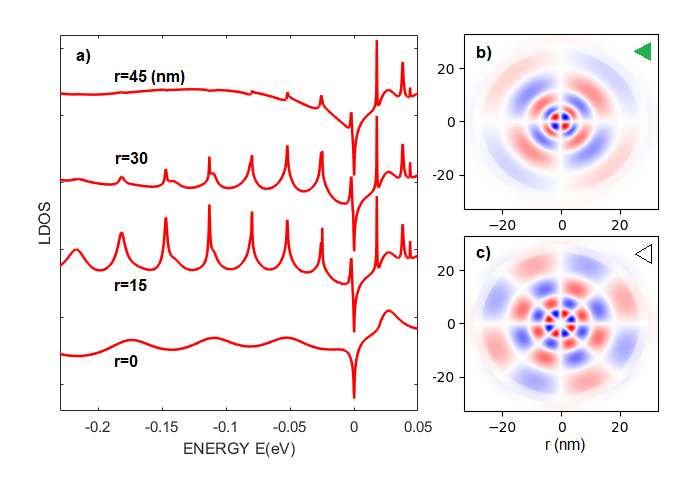}
\end{center}
\caption{\small \textit{(color online) $(a)$ LDOSs($E$) taken from LDOS($E,r$) in Fig.2$(b)$ at different distances $r$ given in the figure; $(b)$ Spatial distributions of the LDOS for the two resonances/QBSs indicated respectively by the arrows in Fig.2$(b)$ [ $j = 1/2 \ (7/2)$ for the lower (higher) state].} } 
\label{fig:3}
\end{figure*}

In Fig.3$(a)$ we specifically display the LDOS($E$), taken from LDOS($E, r$) in Fig.2$(b)$ at different distances $r$, given in the figure. Note that for the CGNPJ-sample studied in this figure the junction-boundary region ranges from $r_i = 15 \ nm$ to $r_f = 45 \ nm$. So, as clearly seen from Fig.3$(a)$, the resonances mainly apprear in the boundary region of the junction. In other words, electronic waves are mainly confined at the junction boundary, manifesting a characteristic feature of  the EWGM-confinement. A similar conclusion can also be deduced from Fig.3$(b)$, where the spatial distributions of the LDOS are plotted for the two resonances/QBSs marked by the corresponding arrows in Fig.2$(b)$ \cite{note02}. The observed ring structure of these distributions is one more manifestation of the EWGM-confinement. Additionally, noting on a difference in the momentum $j$ between these two QBSs, $j = 1/2 (7/2)$ for the lower (higher) level in Fig.2$(b)$, we notice that with increasing $j$ the confinement becomes stronger and the elecronic wave functions become more localised near the junction boundary. This is in full agreement with the ordinary WGM-idea.

Thus, the TDOS in Fig.2$(d)$ indeed shows itself to be a EWGM-spectrum. To quantitativly evaluate this spectrum we measured the resonance energies $E_i$, resonance widths $\delta E_i$ and resonance spacings $\Delta E_i$ for the five most profound resonances labelled by the numbers ($i$ = 1 to 5) in the spectrum. Then we calculated the partial quality-factors $[Q_i ]$ and mode radii $[R_i ]$. Obtained results are as follows: $Q_i (eV) \approx 14.23, \ 34.16, \ 61.40, \ 92.34, 78.40 $ and $R_i (nm) \approx 37.56, \ 36.83, \ 32.81, \ 29.73, \ 29.12 $ as $i$ = 1 to 5. From these data we deduced the average values that characterise the whole EWGM-spectrum in Fig.2$(d)$: the quality factor  $Q \approx 47.37$ and the mode radius, $R \approx 31.83$. Also, the resonance spacing $\Delta E_i$ seems to slightly increase from $27 \ meV$ to $34 \ meV$ as $i$ increases from 1 to 5 with the average value of $\Delta  \approx 31 \ meV$. These obtained values of the mode-radius $R \approx 32 \ nm$ and the resonance spacing $\Delta  \approx 31 \ meV $ seem to be rather reasonable in relation to the junction size (average radius $L = 30 \ nm$). Here, as a reference, we would like  to mention that the values $ R \approx 50 \ nm$ and $\Delta \approx 40 \ meV$ have been reported for the experimental data from Fig.2A in Ref. \cite{zhao2015}. Concerning the $Q$-factor, however, the obtained value $Q \approx 45$ shows a complete surprise, it is too small compared to $Q$-factors in ordinary optical WGM-microresonators ($\approx 10^5 - 10^8$ \cite{vahala2003, matsko2006}). Regretfully, the $Q$-factor is not claimed in Ref.\cite{zhao2015} as well as in the other experiment, relating to the EWGMs in CGNPJs \cite{jiang2017}. So, we tried ourselves to get some rough estimations from the data published. Analysing the three most profound resonances  labelled $1'$, $2'$, and $3'$ from Fig.2D in Ref.\cite{zhao2015} as well as the most profound resonances from Fig.3g in Ref.\cite{jiang2017}, using the same way as described above, we learn that for all these experimental resonances the partial $Q$-factors are in the same order of value as the computing $Q$-factors presented above.       

\begin{figure*}[htb]
\begin{center}
\includegraphics[width=0.8\textwidth]{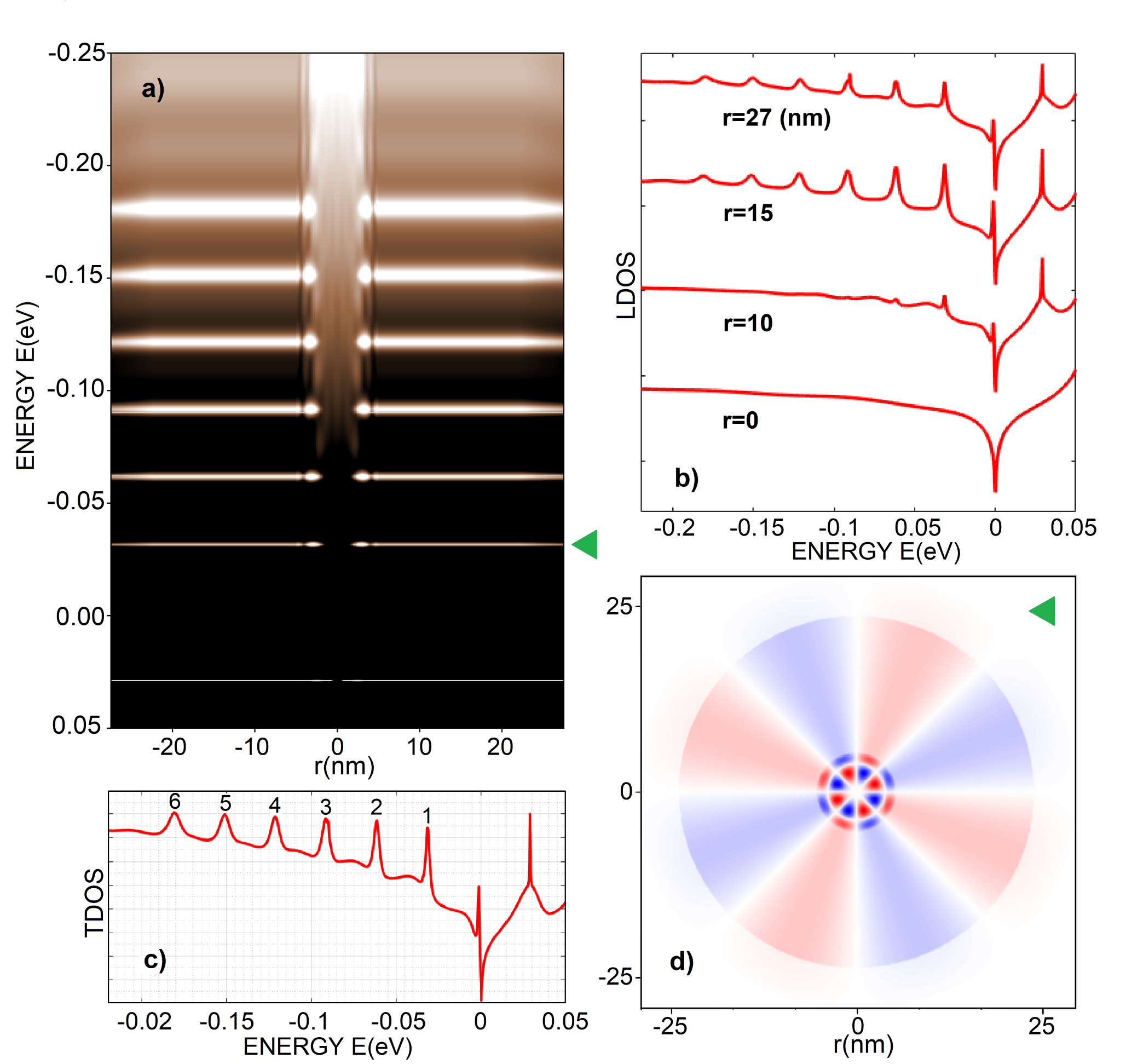}
\end{center}
\caption{\small \textit{(color online) $(a)$, $(b)$, $(c)$, and $(d)$ are respectively the same as Fig.2$(b)$, Fig.3$(a)$, Fig.2$(d)$ and Fig.3$(b)$, but for the CGPP'J with potential parameters: $[ U_l , \ U_g , \ L, \ \alpha ] = [0.28 \ eV, \ 0.06 \ eV, \ 25 \ nm, \ 0.0 ]$ - step junction-boundary potential. }}
\label{fig:4}
\end{figure*}

Next, we present in Fig.4 and Fig.5 the computing spectra obtained for a CGPP'J and CGPNPHJ, respectively. It should be again noted that each of Figs.4 and 5 is very similar in both content and structure to Figs.2$(b)$-$(d)$ plus Fig.3. So, we would like immediately to remark that, like Figs.2-3 for the CGNPJ, Fig.4 (or Fig.5) qualitatively demonstrates an emergence of EWGMs in the CGPP'J (or CGPNPHJ) under study. Note that in accord with the experimental $pp'$-junction measured in Ref.\cite{ren2019} we have chosen the particular sample with a step junction-boundary potential for the first attempt to study EWGMs in CGPP'Js (in Fig.1$(a)$, step junction-boundary potentials are described by the dashed/solid lines with $x_i \equiv x_f$). And, this is the case reported in Fig.4 (with potential parameters given in the figure).  

\begin{figure*}[htb]
\begin{center}
\includegraphics[width=0.8\textwidth]{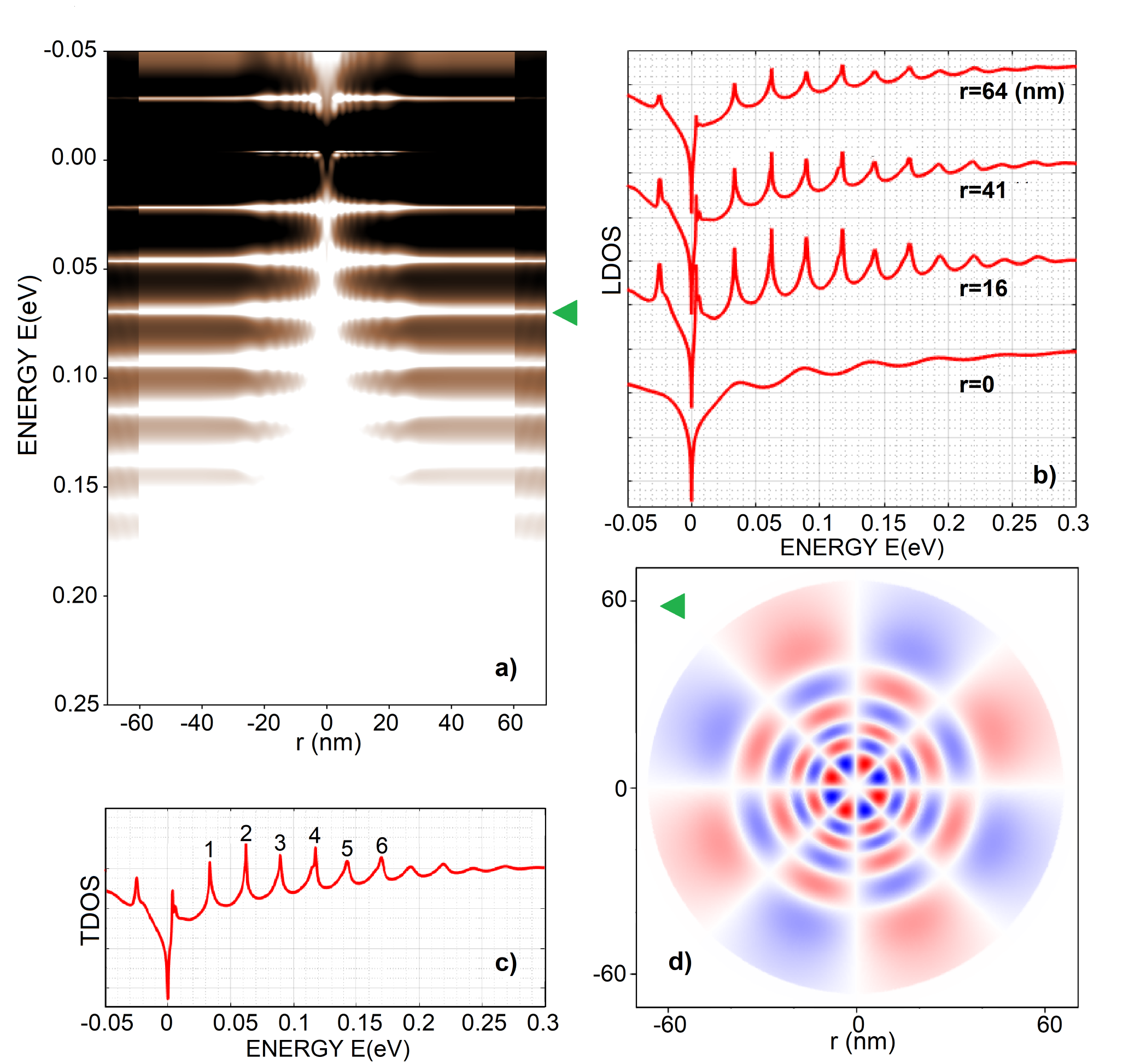}
\end{center}
\caption{\small \textit{(color online)  $(a)$, $(b)$, $(c)$, and $(d)$ are respectively the same as Fig.2$(b)$, Fig.3$(a)$, Fig.2$(d)$ and Fig.3$(b)$, but for the CGPNPHJ with potential parameters: $[U_l , \ U_g , \ L, \ D, \ S ] = [-0.60 \ eV, \ 0.15 \ eV, \ 40 \ nm, \ 25 \ nm, \ 2 \ nm ]$.  }}
\label{fig:5}
\end{figure*}

Quantitatively, analyzing the six most profound resonances labelled by the numbers from 1 to 6 in the  TDOS in Fig.4$(c)$, we obtained for the studied CGPP'J the partial $Q$-factors and mode radii as follows: $(1) \  Q_i \approx 16.09, \ 45.39, \ 18.18, \ 17.49, \ 16.56$ and $15.74$ and $(2) \ R_i (nm) \approx 32.49, \ 33.30, \ 33.86, \ 33.58, \ 33.72$ and $33.86$ as $i = 1$ to $6$. So, on the whole, the studied CGPP'J is characterised by the average quality-factor of $Q \approx 21.57$ and mode radius of $R \approx 33.47 \ nm$. Correspondingly, for the resonance spacing that slightly decreases from $31$ to  $29.7 \ meV$ we have the average value $\Delta \approx 30 \ meV$. Analogously, for the six resonances numbered in the TDOS presented in Fig.5$(c)$ we obtained for the studied CGPNPHJ (as $i = 1$ to $6$): $(1) \ Q_i \approx  21.38, \ 48.53, \ 40.23, \ 82.10, \ 54.64$ and $67.18$ with the average value $Q \approx 46.47$; $(2) \ R_i (nm) \approx  39.92, \ 39.64, \ 38.31, \ 37.56, \ 36.36$ and $35.90$ with the average mode radius $R \approx 37.51 \ nm$; and $(3)$ The average resonance spacing $\Delta  \approx 27 \ meV$. Overall, obtained values of $R$ and $\Delta$ are rather reasonable in relation to the potential parameters of the studied junctions. We would here mention that for the CGPP'J measured in the experiment \cite{ren2019} the average level spacing was reported to be $48 \ meV$. Concerning the $Q$-factors, however, the values obtained for both the CGPP'J in Fig.4 and CGPNPHJ in Fig.5 are very small, in the same order of value as those for CGNPJs analysed in Figs.2-3 . 
            
Thus, it seems that all the three EWGM-spectra presented in Figs.2-5 for circular graphene resonators of different junction-types show very small values of their $Q$-factors. A question may then be arisen about if such the small $Q$-factors are a particular property of the junctions studied. So, we largely searched for  EWGMs, varying parameter values of the potential $U(r)$ for each junction type. As a brief summary, we present in Fig.6 the TDOSs with EWGMs for three resonators of each junction type: $(a)$ CGNPJs; $(b)$  CGPP'Js, and $(c)$ CGPNPHJs (with potential-parameter values given in the figure). Obviously, all these TDOSs show the EWGMs, similar to the TDOSs in Figs.2$(a)$, 4$(c)$ and $5(c)$. Note that some of these TDOSs are specially collected from the junctions with step junction-boundary potential (in the case of CGPNPHJs, it means $x_i \equiv x_m$ and $x_n \equiv x_f$, see Fig.1$(b)$)  

\begin{figure*}[htb]
\begin{center}
\includegraphics[width=0.8\textwidth]{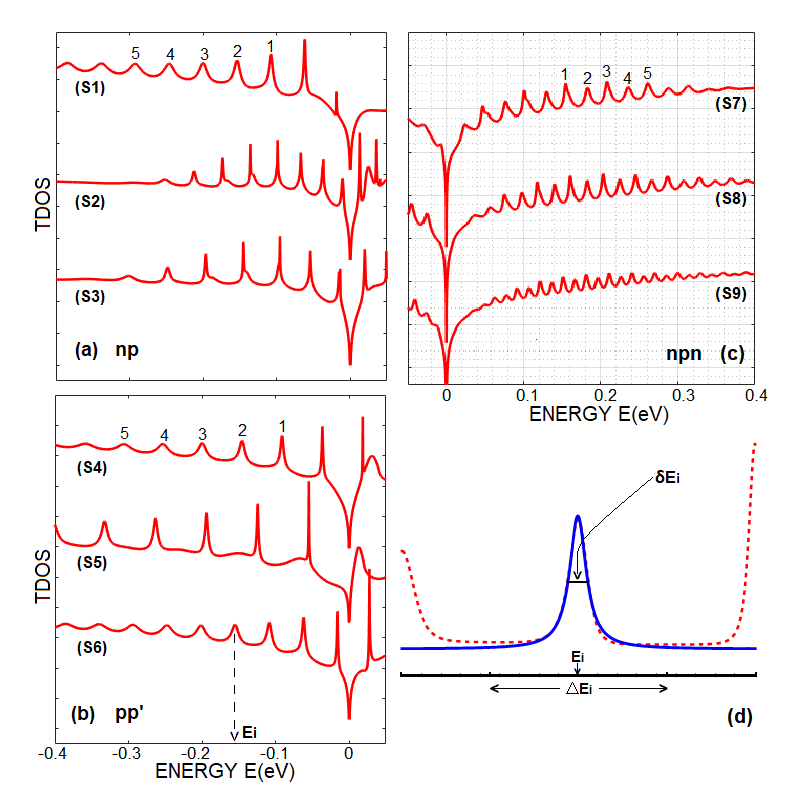}
\end{center}
\caption{\small \textit{(color online) The TDOSs are shown for: $(a)$ three CGNPJs with potential parameters $[U_l (eV) , \ U_g (eV), \ L(nm), \ \alpha ] = [ 0.15, \ -0.02, \ 15, \ 0.0], [ -0.4, \ 0.02, \ 27, \ 0.5]$ and $ [ -0.5, \ 0.05, \ 20, \ 0.5]$;  $(b)$ three CGPP'Js with potential parameters $[U_l (eV) , \ U_g (eV), \ L(nm), \ \alpha ] = [ 0.28, \ 0.04, \ 13, \ 0.2], [ 0.35, \ 0.02, \ 10, \ 0.0]$ and $ [ 0.2, \ 0.02, \ 15, \ 0.1]$; and $(c)$ three CGPNPHJs with potential parameters $[U_l (eV) , \ U_g (eV), \ L(nm), \ D(nm), \ S(nm) ] = [ -0.6, \ 0.15, \ 40, \ 8, \ 8], [ -0.7, \ 0.3, \ 50, \ 15, \ 2]$ and $ [ -0.7, \ 0.3, \ 70, \ 15, 2]$. $(d)$ demonstration of the method used to evaluate the partial $\Delta E_i$ and $\delta E_i$ for the resonance at energy $E_i$ (indicated by the arrow from the lowest curve in Fig.6$(c)$: the resonance peak (red-dashed) is fitted to the appropriate Lorentzian profile (solid-blue) }}
\label{fig:6}
\end{figure*}

Quantitative analyses of all the EWGM-spectra shown in Fig.6 are in detail given in Tables S1 and S2 (Supplementary Materials). As can be seen in Table S2, the values of mode radii $R$ and resonance spacings $\Delta$ obtained for all the examined resonators, $(s1)$ to $(s9)$ (each with five numbered resonances - see Fig.6$(a-c)$), vary from $\approx 14$ to $\approx 67 \ nm$  and from $\approx 15$ to $\approx 69 \ nm$, respectively. These values of $R$ and $\Delta$ are in the same order of value as the corresponding data reported in Figs.2-5 and seem rather reasonable, depending on the resonator size. As for the quality factors, though the three CGPNPHJs, $(s7)$ to $(s9)$, show somewhat improved values of $Q$, about a hundred, totally, for all examined resonators, the $Q$-factors are still small, $\leq 10^2$.   
We would like here to emphasise that such the $Q$-factors are found in the EWGMs emerged in all the electrostatic-potential induced circular graphene junctions under study, regardless of the junction type as well as the smoothness of junction-boundary potentials.

Lastly, we would clarify in Fig.6$(c)$ the way we have used to evaluate the EWGM-characteristics. For the resonance (or QBS) of interest (for instance, the resonance marked by the arrow in the last curve in Fig.6$(b)$) the quantities to be determined are as follows: $(i)$ the resonance energy $E_i$ that appears as the eigenvalue of the Dirac equation, $(ii)$ the resonance width $\delta E_i$ that is determined by fitting the resonant peak (dashed line) to an appropriate Lorentzian profile (solid line), following the standard way of evaluating this quantity (see, for example, Ref.\cite{davies1998}), and the resonance spacing $\Delta E_i$ that is determined as shown in Fig.6$(c)$. The quantities $E_i$ and $\delta E_i$ are then used to evaluate partial $Q_i$ and $R_i$ as described above. 

In conclusion, we have theoretically studied the EWGMs emerged in energy spectra of electrostatic-potential induced circular graphene junctions, including all types of uni-junctions as well as bipolar-junctions. To this end, we modelled the studied junctions by appropriate electrostatic confinement potentials and calculated the LDOSs of structures within the framework of the Dirac-Weyl formalism for massless fermions. Calculations have been carried out for many junction-samples of each junction-type, varying potential-parameter values. From obtained LDOSs we identified those with EWGMs, following the way of identifying the optical WGMs. It seems that EWGMs could be emerged in energy spectra of circular graphene resonators with any junction-type, uni-junctions or bipolar junctions, including those with a step junction-boundary potential. For all the identified EWGMs we evaluted their characteristics such as the $Q$-factors, mode radii $R$, and resonance spacings $\Delta$. Obtained values of $R$ and $\Delta$ are rather reasonable, depending on the potential parameters. However, the $Q$-factors seem always to be surprisingly small (generally, $\leq 10^2$). These theoretical results, including small $Q$-factors, describe rather well the existent experimental data. We assume that an observed smallness of $Q$-factors is mainly due to the Klein tunnelling. On the one hand, the Klein tunnelling creates the resonances/QBSs in EWGM-spectra. On the other hand, Klein tunnelling itself diminishes the resonance/QBS life-times and, therefore, the $Q$-factors of these resonances. In this view, we would speculate that a smallness of $Q$-factors is a general property of all graphene resonators created by electrostatic potentials, regardless of resonator shape and size. A magnetic field might enhance a localisation, but it also induces weak resonances, destroying the WGM-feature of spectra \cite{ren2019}.

{\sl Acknowledgments}. We are very grateful to H.Chau Nguyen for helpful discussions. We also thank T.T. Nhung Nguyen, T.D. Linh Dinh and H. Minh Lam for some collaboration in the first step of numerical computations. This work is funded by Vietnam National Foundation for Science and Technology Development (NAFOSTED) under Grant No. 103.02-2015.48.

\end{document}